\pdfoutput=1 
\documentclass{JINST}

\title{New approaches for boosting to uniformity}

\author{
Alex Rogozhnikov$^{a,b}$\thanks{Corresponding author.}~,
Aleksandar Bukva$^c$, 
Vladimir Gligorov$^d$,
Andrey Ustyuzhanin$^{b,e,f}$ and
Mike Williams$^g$\\
\llap{$^a$}Lomonosov Moscow State University, Moscow, Russia\\
\llap{$^b$}Yandex, Moscow, Russia\\
\llap{$^c$}Faculty of Physics, Belgrade, Serbia \\
\llap{$^d$}Organisation Europ\'eenne pour la Recherche Nucl\'eaire (CERN), Geneva, Switzerland  \\
\llap{$^e$}Moscow Institute of Physics and Technology, Moscow, Russia\\
\llap{$^f$}Imperial College, London, UK\\
\llap{$^g$}Massachusetts Institute of Technology, Cambridge, MA, United States \\
E-mail: \email{alex.rogozhnikov@yandex.ru}}

\abstract{
The use of multivariate classifiers has become commonplace in particle physics.
To enhance the performance, a series of classifiers is typically trained; this is a technique known as boosting. 
This paper explores several novel boosting methods that have been designed to produce a uniform selection efficiency in a chosen multivariate space.  
Such algorithms have a wide range of applications in particle physics,
from producing uniform signal selection efficiency across a Dalitz-plot to avoiding the creation of false signal peaks in an invariant mass distribution when searching for new particles.

}

\usepackage{amsmath, amsthm, amsfonts, xspace}
\usepackage{caption}
\usepackage{subcaption}
\usepackage{float}	

\theoremstyle{definition}

\theoremstyle{remark}

\newcommand{\abs}[1]{\left\vert#1\right\vert}

\def\x {\ensuremath{\vec{x}}\xspace}
\def\y {\ensuremath{\vec{y}}\xspace}
\def\knn{\text{knn}}
\def\knni{\text{knn}(i)}

\begin{document}
\maketitle

\section{Introduction}

Methods of machine learning play an important role in modern particles physics. 
Multivariate classifiers, {\em e.g.}, boosted decision trees (BDTs) and artificial neural networks (ANNs), are now commonly used in analysis selection criteria. BDTs are now even used in software triggers~\cite{ref:lhcbhlt,ref:bbdt}. 
To enhance the performance, a series of classifiers is typically trained; this is a technique known as boosting.
Boosting  involves training many simple classifiers and then building a single composite classifier from their responses.
The classifiers are trained in series with the inputs of each member being augmented based on the performance of its predecessors.  This augmentation is designed such that each new classifier targets those events which were poorly classified by previous members of the series.  The classifier obtained by combining all members of the series is typically much more powerful than any of the individual members. 

In particle physics, the most common usage of BDTs is in classifying candidates as signal or background.  The BDT is determined by optimizing some figure of merit (FOM), {\em e.g.}, the signal purity or approximate signal significance.  This approach is optimal for a counting experiment; however, in many cases the BDT-based selection obtained in this way is not optimal.  
For example, in a Dalitz-plot (or any angular or amplitude analysis) analysis, obtaining a selection efficiency for signal candidates that is uniform across the Dalitz-plot is more important than any integrated FOM.  Similarly, when measuring a mean particle lifetime, obtaining an efficiency that is uniform in lifetime is what is desired.  In both cases, obtaining a uniform selection efficiency greatly reduces the systematic uncertainties involved in the measurement.  
When searching for a new particle, an analyst may want a uniform efficiency in mass for selecting background candidates so that the BDT-based selection does not generate a fake signal peak.  Furthermore, the analyst may also desire a uniform selection efficiency of signal candidates in mass (or other variates) since the mass of the new particle is not known.  In such cases, the BDT is often trained on simulated data generated with several values of mass (lifetime, {\em etc.}).  A uniform selection efficiency in mass ensures that the BDT is sensitive to the full range of masses involved in the search.

\section{Uniformity Boosting Methods}
The variates used in the BDT are denoted by \x, while the variates in which uniformity is desired are denoted by \y.  Some (perhaps all) of the \x variates will be {\em biasing} in \y, {\em i.e.} they provide discriminating power between signal and background that varies in \y.  A uniform BDT selection efficiency can be obtained by removing all such variates; however, this will also reduce the power of the BDT.  The goal of boosting algorithms presented in this paper is to {\em balance} the biases to produce the optimal uniform selection.

One category of boosting works by assigning training events more weight based on classification errors made by previous members of the series.  For example, the AdaBoost~\cite{ref:FS1997} algorithm updates the weight of event $i$, $w_i$, according to
\begin{equation}
w_i^{\prime} = w_i \times {\rm exp}\left[-\gamma_i p_i\right], 
\end{equation}
where $\gamma = +1(-1)$ for signal(background) events and $p$ is the prediction for each event produced by last classifier in the series.  
The uBoost technique, described in detail in Ref.~\cite{ref:uboost}, alters the event-weight updating procedure to achieve uniformity in the signal-selection efficiency.

Another approach to obtain uniformity, introduced in this paper, involves defining a more general expression of the AdaBoost criteria:
\begin{equation}
 w_i^{\prime} = w_i \times {\rm exp}\left[-\gamma_i \sum_j a_{ij} p_j\right],
\end{equation}
where $a_{ij}$ are the elements of some square matrix $A$.  For the case where $A$ is the identity matrix, the AdaBoost weighting procedure is recovered.  
Other choices of $A$ will induce non-local effects, {\em e.g.}, consider the sparse matrix $A_{\rm knn}$ given by
\begin{equation}
  a_{ij}^{\rm knn} = \begin{cases} 
    \frac{1}{k}, & j \in \knni, \text{ events $i$ and $j$ belong to the same class} \\
    0, & \text{otherwise},
\end{cases}
\end{equation}
where $\knn(i)$ denotes the set of $k$-nearest-neighbor events to event $i$.
This procedure for updating the event weights, which we refer to as kNNAdaBoost, accounts for the score of each event's $k$ nearest neighbors and not just each event individually.

The gradient boosting~\cite{ref:F1999} (GB) algorithm category requires the analyst to choose a differentiable loss function with the goal of building a classifier that minimizes the loss.  
A popular choice of loss function is the so-called AdaLoss function 
\begin{equation}
L_{\rm ada} = \sum\limits_{i=1}^{n} {\rm exp}\left[-\gamma_i s_i\right]. 
\end{equation}
The {\em scores} $s$ are obtained for each event as the sum of predictions of all elements in the series. 
At each stage in the gradient boosting process, a {\em regressor} (a decision tree in our case) is trained whose purpose is to decrease the loss.  This is accomplished using the gradient decent method and the {\em pseudo-residuals}
\begin{equation}
  -\frac{\partial L_{\rm ada}}{\partial s_i} = \gamma_i  {\rm exp}\left[-\gamma_i s_i\right],
\end{equation}
which are positive(negative) for signal(background) events and have larger moduli for poorly classified events.  

The gradient-boosting algorithm is general in that it only requires the analyst specify a loss function and its gradient.  The AdaLoss function considers each event individually, but can easily be modified to take into account non-local properties of the classifier as follows:
\begin{equation}
  \label{eq:loss_general}
  L_{\rm general} = \sum\limits_{i=1}^{n} {\rm exp}\left[-\gamma_i \sum\limits_{j} a_{ij} s_j \right]. 
\end{equation} 
For example, the loss function obtained from Eq.~\ref{eq:loss_general} using\footnote{A natural choice is a square $n\times n$ matrix, but this is not required.} 
$A_{\rm knn}$, which we refer to as kNNAdaLoss and denote $L_{\rm knn}$, accounts for the score of each event's $k$ nearest neighbors and not just each event individually.
The pseudo-residuals of $L_{\rm knn}$ are 
\begin{equation}
  -\frac{\partial L_{\rm knn}}{\partial s_k} = \sum\limits_i \gamma_i a_{ik}^{\rm knn} {\rm exp}\left[-\gamma_i \sum_j a_{ij}^{\rm knn} s_j \right].
\end{equation}
One can see that the direction of the gradient will be influenced the most by events whose $k$-nearest-neighbor events are classified poorly. 
We generically refer to GB methods designed to achieve uniform selection efficiency as uniform GB (uGB).  The specific algorithm that uses kNNAdaLoss will be called uGBkNN.  

Another approach is to include some uniformity metric in the definition of the loss function.  Consider first the case where the data have been binned in \y.  If the distribution of classifier responses in each bin, $f_b(s)$, is the same as the global response distribution, $f(s)$, then any cut made on the response will produce a uniform selection efficiency in \y.  Therefore, performing a one-dimensional goodness-of-fit test of the hypothesis $f_b \equiv f$ in each bin provides an assessment of the selection uniformity.  
For example, one could perform the Kolmogorov-Smirnov test in each bin and define a loss function as follows:
\begin{equation}
  L_{\rm flat(KS)} = \sum\limits_{b} w_b {\rm max}|F_b(s)-F(s)|,
\end{equation}
where $F_{(b)}(s)$ denotes the cumulative distribution of $f_{(b)}(s)$ and $w_b = \sum \delta({\rm bin}_i - b)/n_{\rm signal}$, {\em i.e.} $w_b$ is the fraction of signal events in the bin\footnote{If weighted events are used, then the fractional sum of weights should be used for $w_b$.}.   

The gradient of the Kolmogorov-Smirnov loss function is zero for events with responses greater than the value of $s$ at which ${\rm max}|F_b(s)-F(s)|$ occurs.  Thus, it is not suitable for gradient boosting due to its instability.  Instead, we use the following {\em flatness} loss function:
\begin{equation}
  L_{\rm flat} = \sum\limits_{b} w_b \int |F_b(s)-F(s)|^2 ds,
\end{equation}
whose pseudo-residuals are ($b$ is the bin containing the $k$th event)
\begin{equation}
 -\frac{\partial L_{\rm flat}}{\partial s_k} \approx - 2 w_b \left[F_b(s_k)-F(s_k)\right] .
\end{equation}
This so-called flatness loss penalizes non-uniformity but does not consider the quality of the classification.  Therefore, the full loss function used is 
\begin{equation}
\label{eq:L_ada_flat}
  L_{{\rm ada}+{\rm flat}} = L_{\rm flat} + \alpha L_{\rm ada},
\end{equation}  
where $\alpha$ is a real-valued parameter that is typically chosen to be small.  The first term in Eq.~\ref{eq:L_ada_flat} penalizes non-uniformity, while the second term penalizes poor classification.  
We refer to this algorithm as uGB with flatness loss (uGBFL). 
In principle, many different flatness loss functions can be defined and could be substituted for our choice here.  See Appendix~A for a detailed discussion on this topic. 

The loss function given in Eq.~\ref{eq:L_ada_flat} can also be constructed without binning the data using k-nearest-neighbor events.  
The cumulative distribution $F_{\rm knn}(s)$ is easily obtained and the bin weight, $w_b$, is replaced by a k-nearest-neighbor weight, $w_{\rm knn}$.  First, each event is weighted by the inverse of the number of times it is included in the k-nearest-neighbor sample of another event.  Then, $w_{\rm knn}$ is the sum of such weights in a k-nearest-neighbor sample divided by the total sum of such weights in the full sample.   This procedure is followed to offset the fact that some events are found in more k-nearest-neighbor samples than other events.  
We study two versions of uGBFL below: uGBFL using bins denoted by uGBFL(bin) and uGBFL using kNN collections denoted by uGBFL(kNN).  
The algorithms are summarized in Table~\ref{tab:algs}.

\begin{table}
  \begin{center}
    \caption{\label{tab:algs} Description of uniform boosting algorithms.}
    \begin{tabular}{c|c}
      \hline
      Name & Description \\
      \hline
        uBoost & algorithm introduced in Ref.~\cite{ref:uboost}\\
      \hline
        kNNAda & AdaBoost modification using matrix $A_{\rm knn}$ \\
        uGBkNN & gradient boost using kNNAdaLoss loss function \\
        uGBFL(bin) & gradient boost using flatness loss $+ \alpha$ AdaLoss as in Eq.~\ref{eq:L_ada_flat} (data binned for FL) \\
        uGBFL(kNN) & same as uGBFL(bin) except kNN events are used rather than bins \\
      \hline
    \end{tabular}
  \end{center}
\end{table}

\section{Example Analysis}


The example analysis studied here involves a so-called Daltiz-plot analysis.  In such analyses, the distribution of events in a 2-D space is typically fit to extract some information of physical interest.  The regions of the Daltiz-plot that tend to have the highest sensitivity to the desired information are the {\em edges}.  Unfortunately, the edge regions also typically have the most background contamination and the least discrimination against background.  Therefore, traditional classifier-based selections tend to produce selections for Dalitz-plot analyses with lower efficiency near the edges.   

This study uses simulated event samples produced using the official LHCb simulation framework.
The software used for the generation of the events is described in LHCb publications as follows :

\begin{quote}
In the simulation, $pp$ collisions are generated using
PYTHIA~\cite{Sjostrand:2006za} 
with a specific LHCb configuration~\cite{LHCb-PROC-2010-056}.  Decays of hadronic particles
are described by EvtGen~\cite{Lange:2001uf}, in which final state
radiation is generated using PHOTOS~\cite{Golonka:2005pn}. The
interaction of the generated particles with the detector and its
response are implemented using the GEANT
toolkit~\cite{Allison:2006ve, Agostinelli:2002hh} as described in
Ref.~\cite{LHCb-PROC-2011-006}.
\end{quote}

All simulated event samples are generated inside the LHCb detector acceptance.
The signal used in this analysis consists of $D_s^\pm\to\pi^+\pi^-\pi^\pm$ decays, simulated
using the $\textrm{D}\_\textrm{DALITZ}$ model of EvtGen to simulate the intermediate resonances which contribute to the
three pion final state. The background candidates are three pion combinations reconstructed in
simulated samples of $c\bar{c}$ and $b\bar{b}$ events, where the charm and bottom quark decays are
inclusively modelled by EvtGen. The simulated events contain ``truth'' information which identifies them as
signal or background, and which identifies the physical origin of the three pion combinations reconstructed
in the $c\bar{c}$ and $b\bar{b}$ simulated samples.

\begin{figure}[] 
  \centering 
  \includegraphics[width=0.49\textwidth]{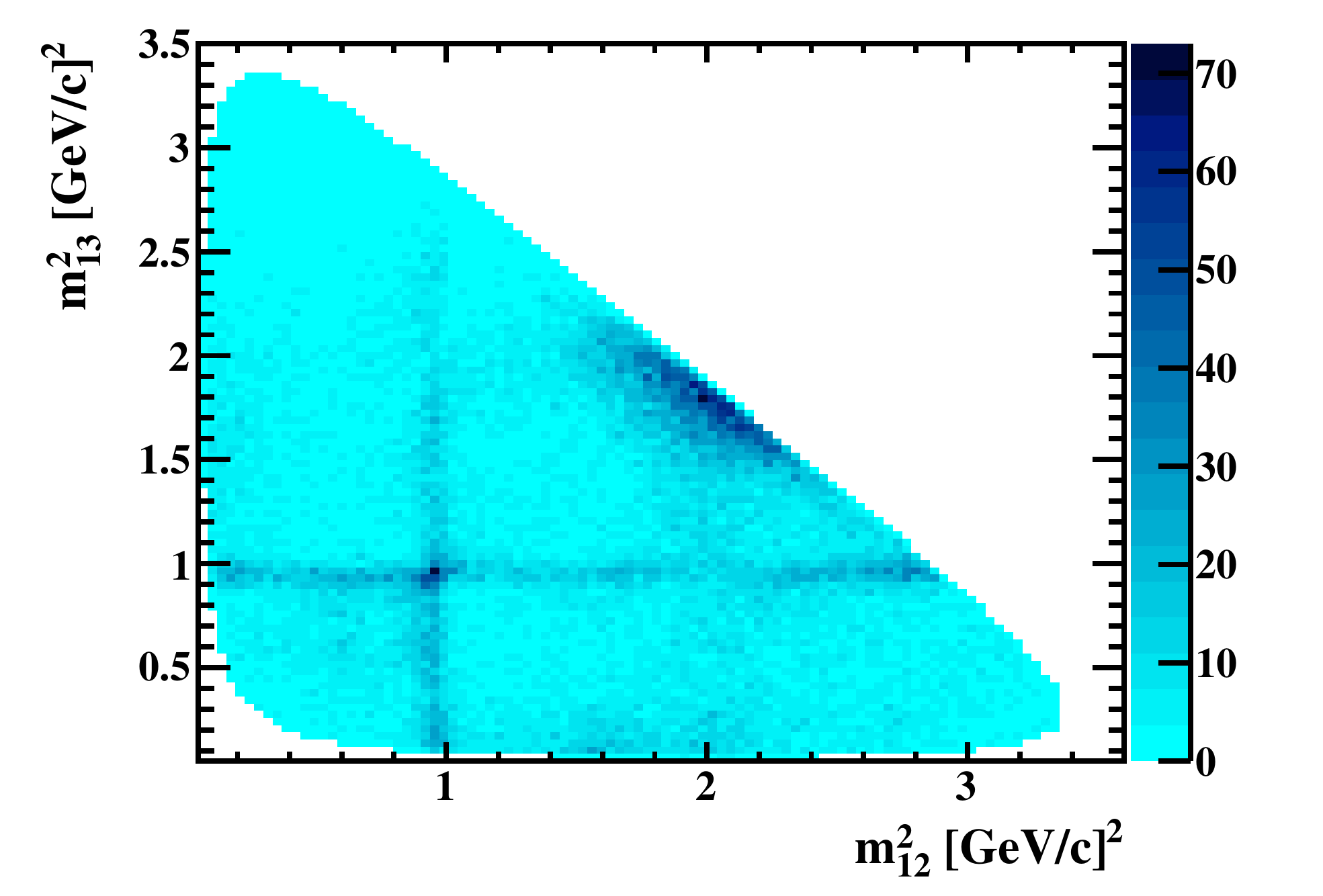}
  \includegraphics[width=0.49\textwidth]{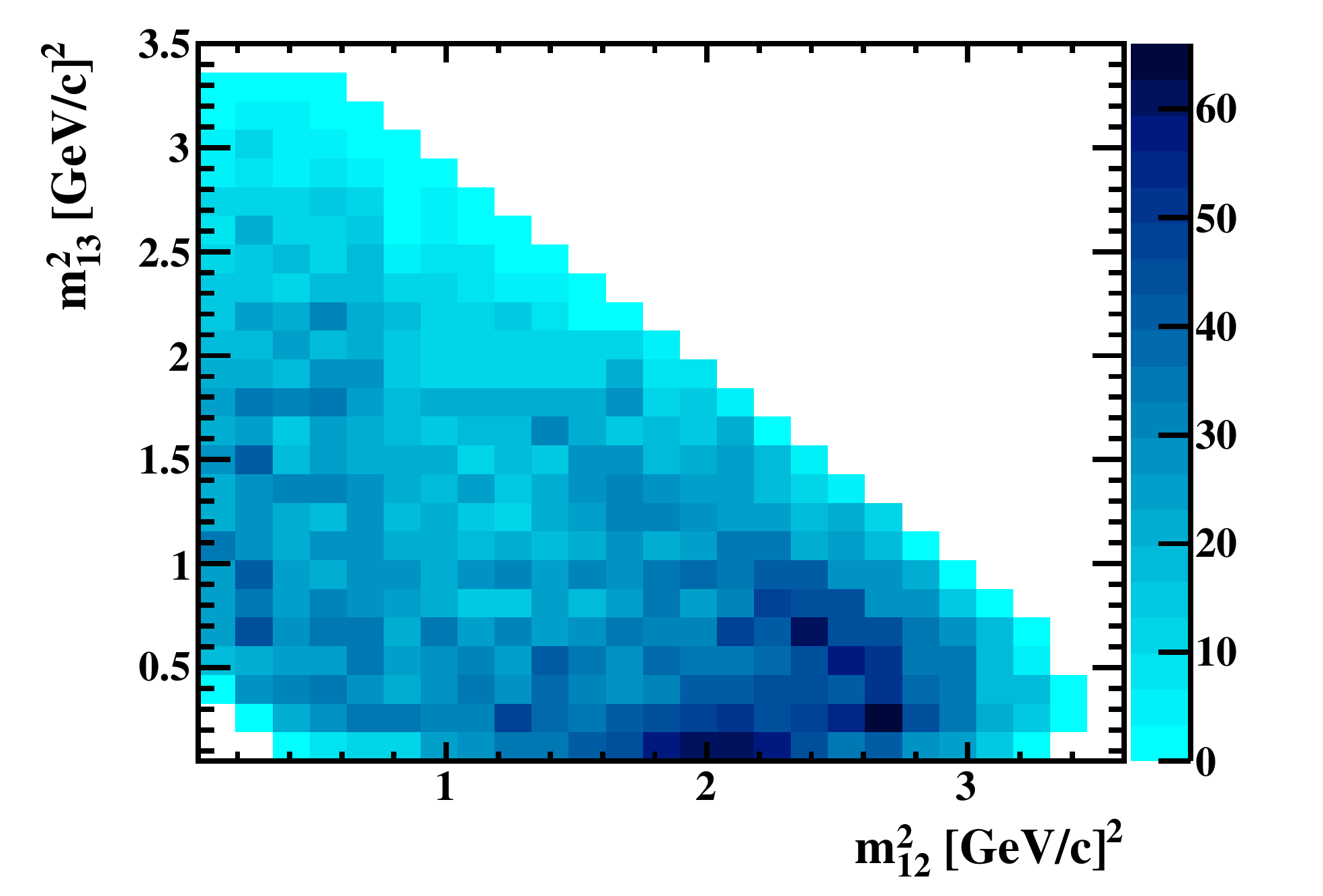}
  \caption{\label{fig:dalitz} Dalitz-plot distributions for (left) signal and (right) background for the $D_s^\pm\to\pi^+\pi^-\pi^\pm$.  The three pions are labeled here as 1, 2 and 3 and ordered according to increases momentum.}
\end{figure}

\begin{figure}[] 
  \centering 
  \includegraphics[width=0.49\textwidth]{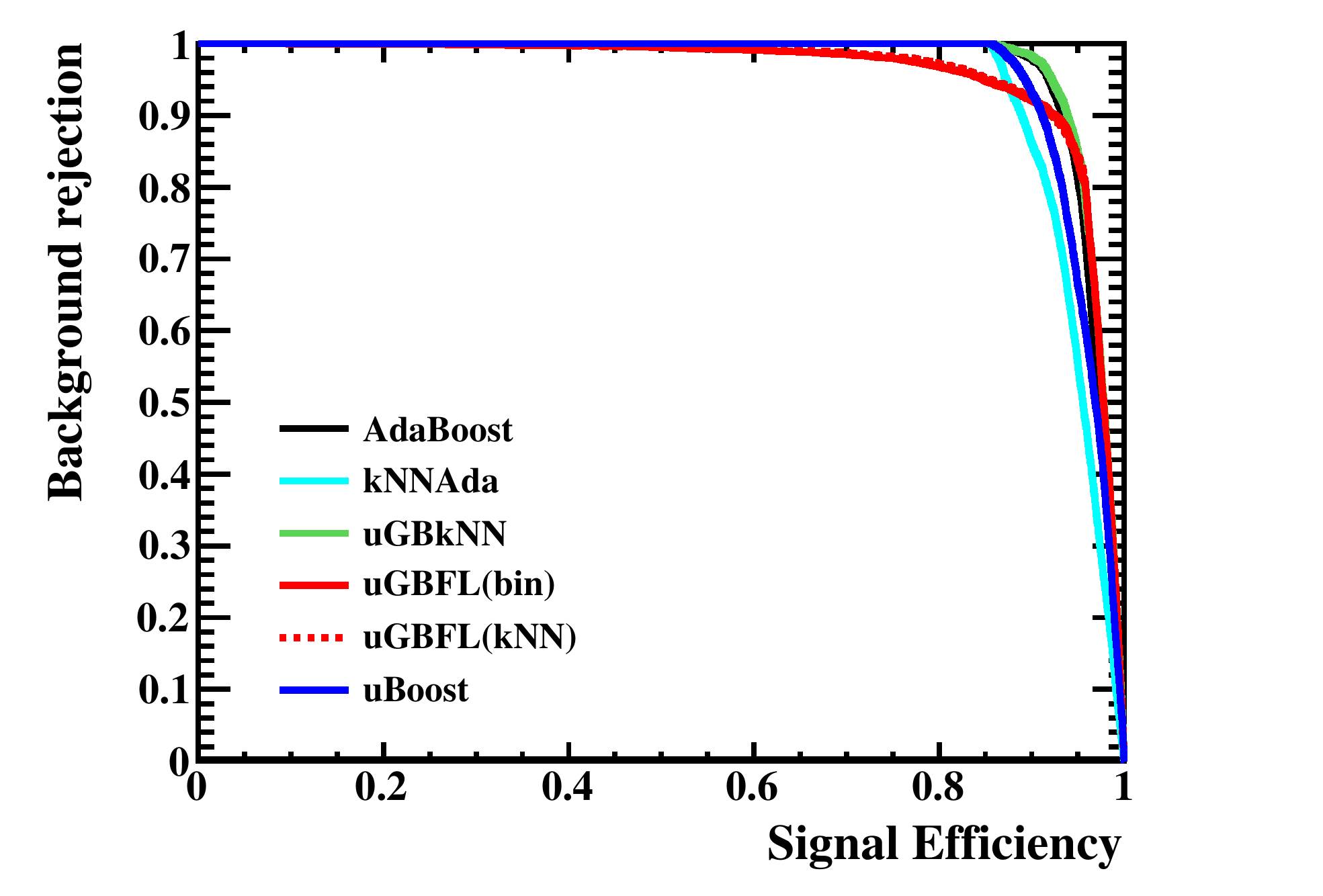}
  \includegraphics[width=0.49\textwidth]{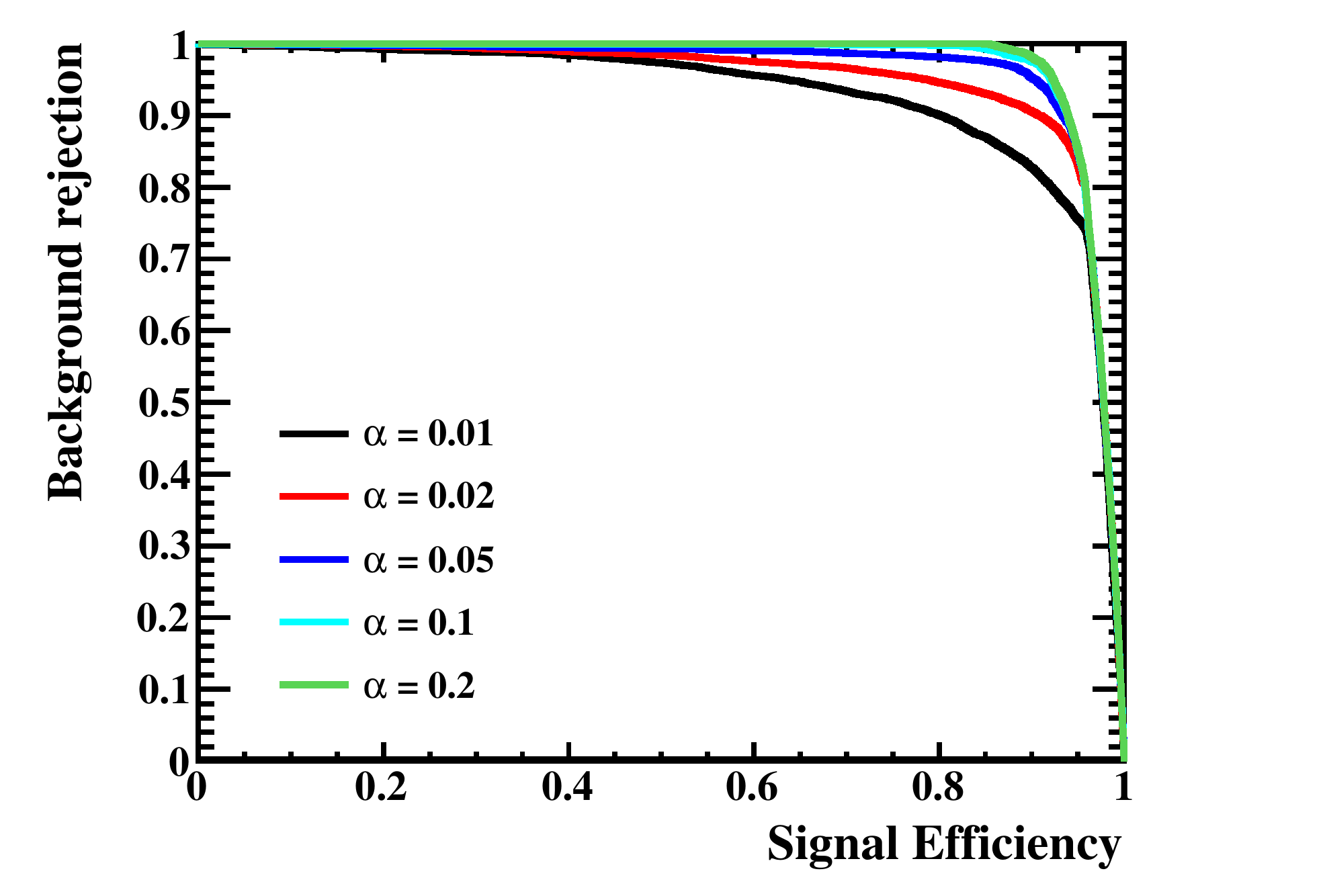}\\
  \caption{\label{fig:dalitz_rocs} (left) ROC curves for classifier algorithms studied in this paper.  For the uGBFL algorithms $\alpha=0.02$ is shown.  (right) ROC curves for uGBFL(bin) for differnet values of $\alpha$.}
\end{figure}

\begin{figure}[] 
  \centering 
  \includegraphics[width=0.49\textwidth]{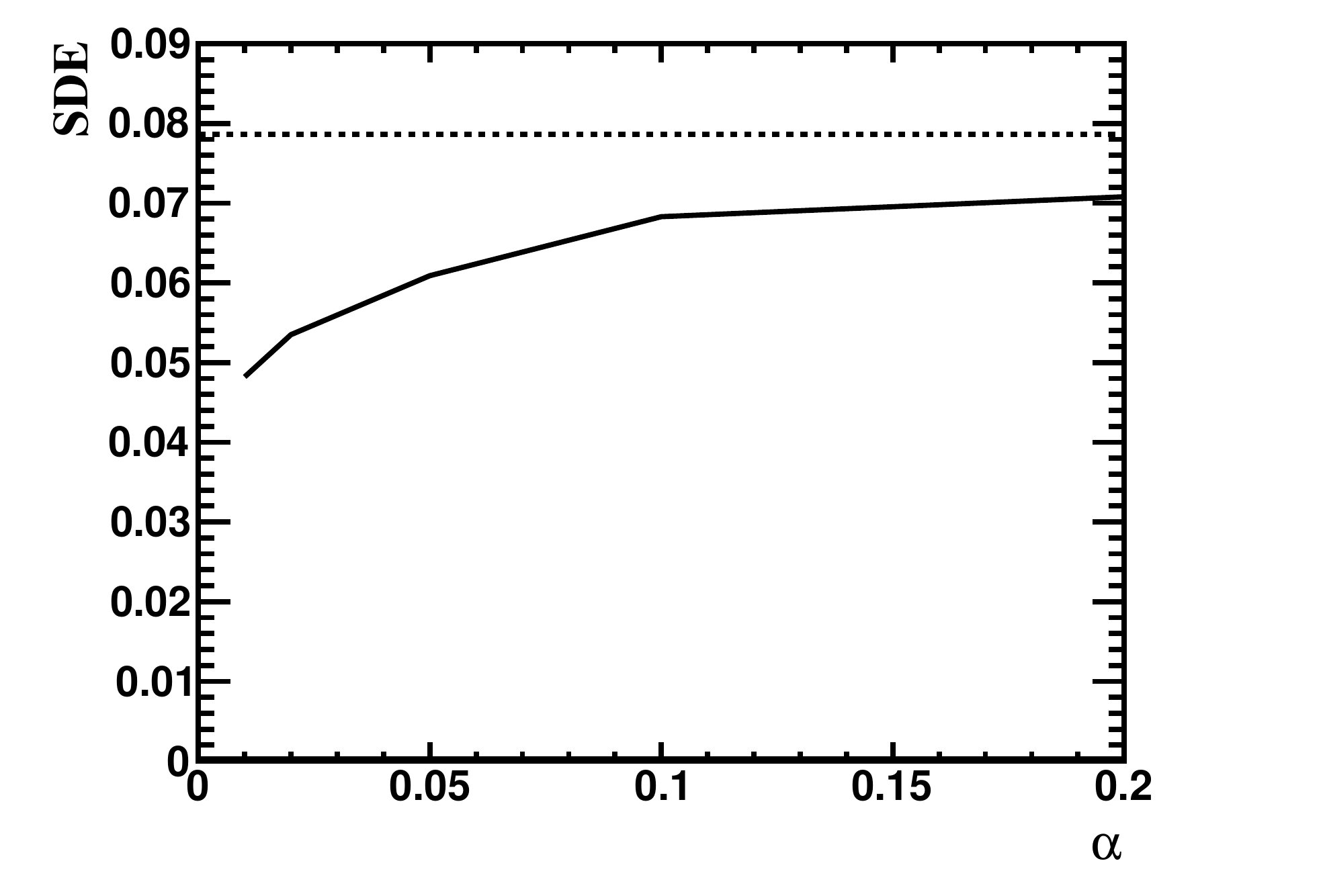}
  \caption{\label{fig:dalitz_sde_v_alpha} Uniformity of the selection efficiency across the Dalitz plot, as measured using the so-called SDE metric described in detail in the appendix, {\em vs} $\alpha$ for uGBFL(bin).  The dashed line indicates the SDE value for AdaBoost.  Lower values of $\alpha$ produce more uniform selection efficiencies.}
\end{figure}

Figure~\ref{fig:dalitz} shows the Dalitz-plot distributions for signal and background events.  These samples are split into training and testing samples and then various BDTs are trained.  For the BDTs designed to produce uniform selections, the $\vec{y}$ variates are the Dalitz masses with the choice of uniform selection efficiency on signal candidates in the Dalitz-plot.  
Figure~\ref{fig:dalitz_rocs} shows the ROC curves obtained for the various classifiers studied in this paper.  For the uGBFL algorithms, there is a choice to be made for the value $\alpha$ which defines the relative weight of the flatness loss {\em vs} AdaLoss.  As expected, increasing $\alpha$, which increases the weight of AdaLoss, drives the ROC curve to be similar to AdaBoost.  Analysts will need to choose how much ROC performance to sacrifice to gain uniformity in the selection efficiency.  In general, the ROC curves for the uniform-driven BDTs are not too different from AdaBoost.  Figure~\ref{fig:dalitz_sde_v_alpha} shows how the uniformity of the selection efficiency depends on $\alpha$.  As expected, as $\alpha$ is decreased the selection becomes more uniform.

\begin{figure}[] 
  \centering 
  \includegraphics[width=0.49\textwidth]{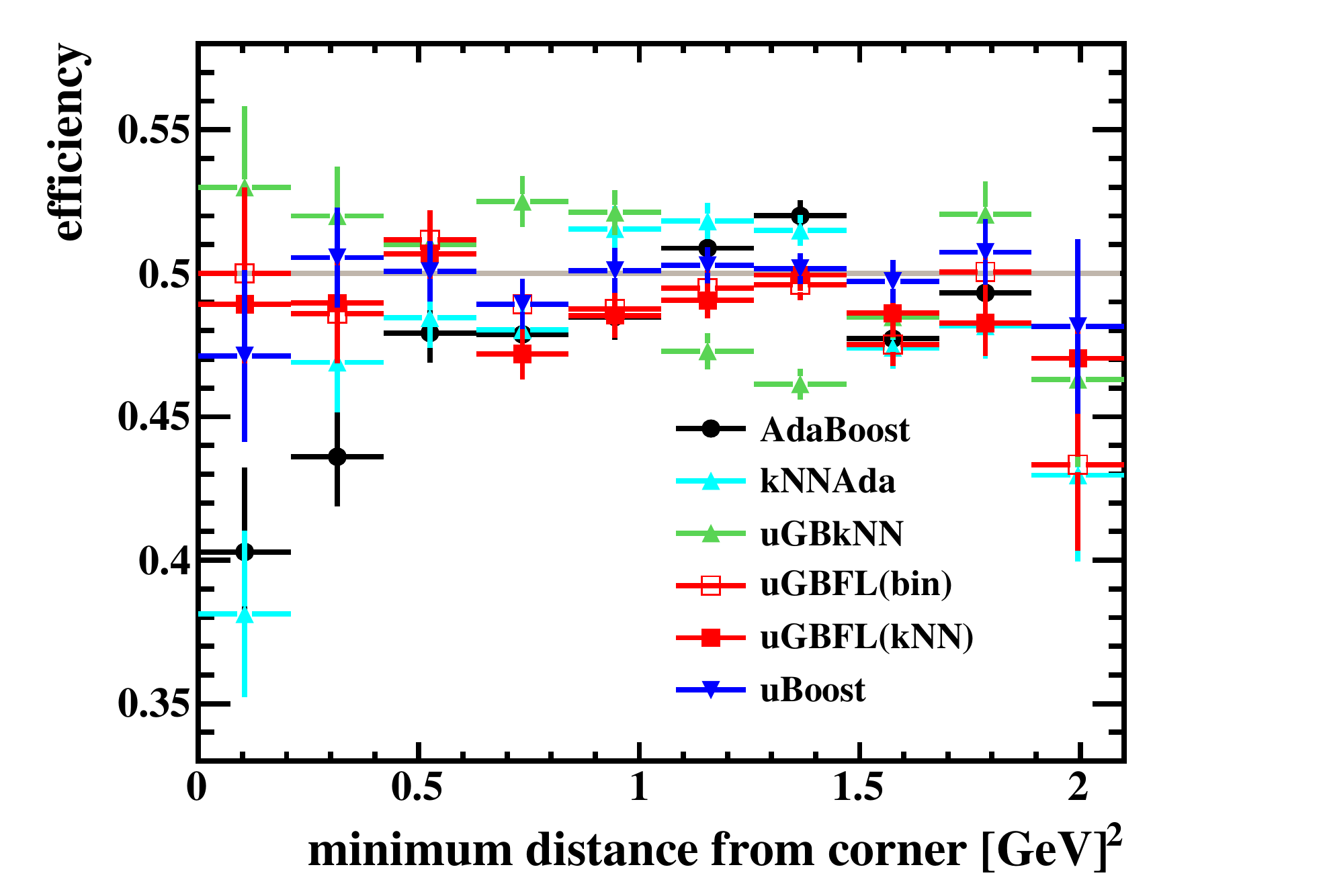}
  \caption{\label{fig:dalitz_results} Efficiency {\em vs} distance to a corner of the Dalitz-plot.  An arbitrary working point of 50\% integrated efficiency is displayed. For the uGBFL algorithms $\alpha=0.02$ is shown.}
\end{figure}

Figure~\ref{fig:dalitz_results} shows the efficiency obtained for each classifier {\em vs} distance from the a corner of the Dalitz-plot\footnote{The Dalitz-plot is essentially a triangle with three corners. Our definition of this distance is ${\rm MIN}\left[(m(D_s)-m(\pi))^2 - m_{ij}^2\right]$, where $ij$ is 12, 13 and 23.}.  The AdaBoost algorithm, as expected, produces a much lower efficiency in the interesting corner regions.  The kNNAdaBoost algorithm does not improve upon the AdaBoost result much.  This is likely due to the fact that while kNNAdaBoost uses non-local kNN information, it does not utilize global information.  The uGBkNN algorithm overcompensates and drives the efficiency higher at the corners.  This suggests that if this algorithm is to be used some stopping criteria or {\em throttle} of the event-weighting updating should be implemented.  
The uGBFL (binned and unbinned kNN) and uBoost algorithms each produce an efficiency which is statistically consistent with uniform across the Dalitz plot.  
As stated above, the analyst is free to optimize the choice of $\alpha$ for uGBFL by defining a metric that involves signal efficiency, background rejection and uniformity, {\em e.g.}, using uniformity metrics discussed in detail in Appendix~A.  

\begin{figure}[] 
  \centering 
  \includegraphics[width=0.49\textwidth]{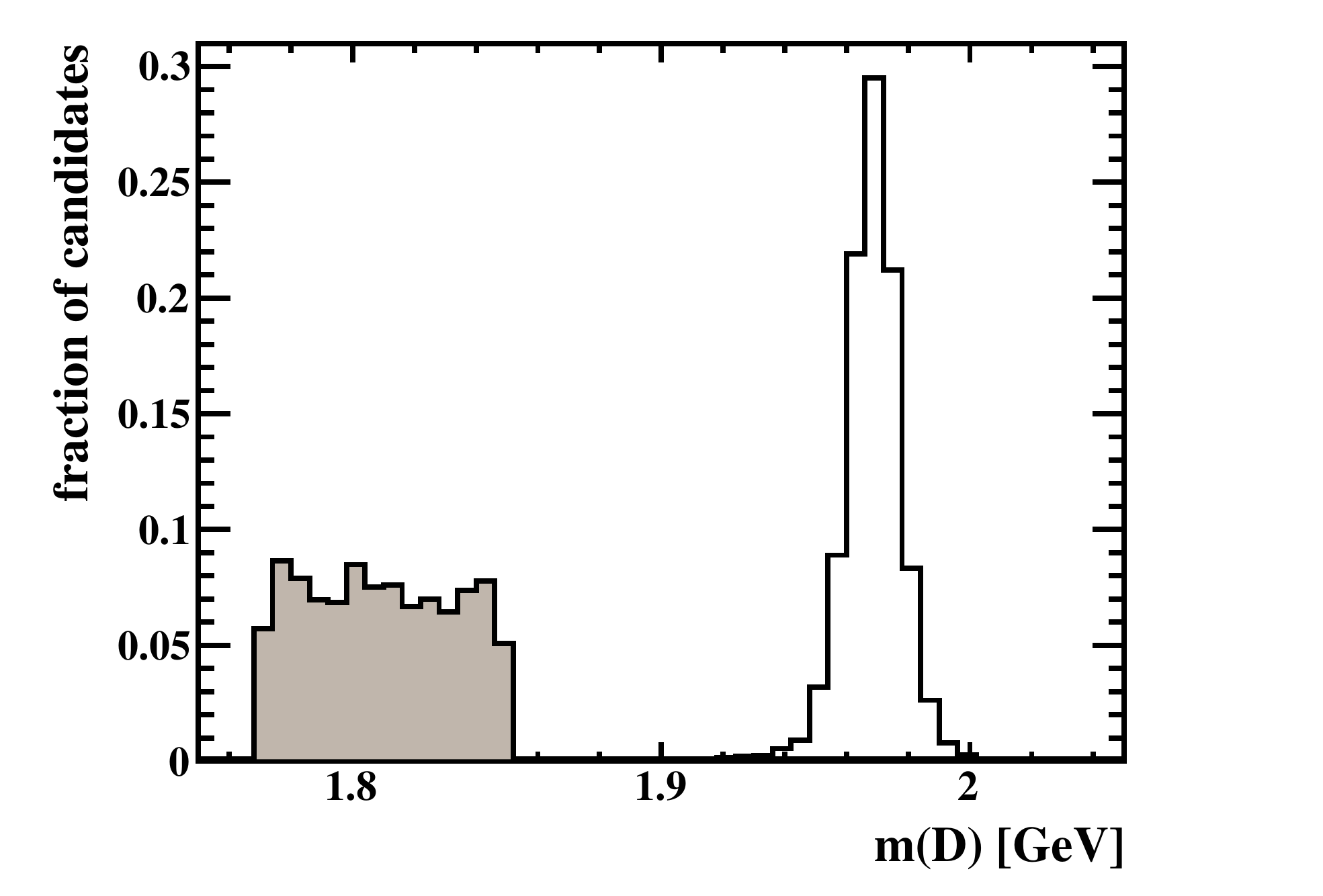}
  \includegraphics[width=0.49\textwidth]{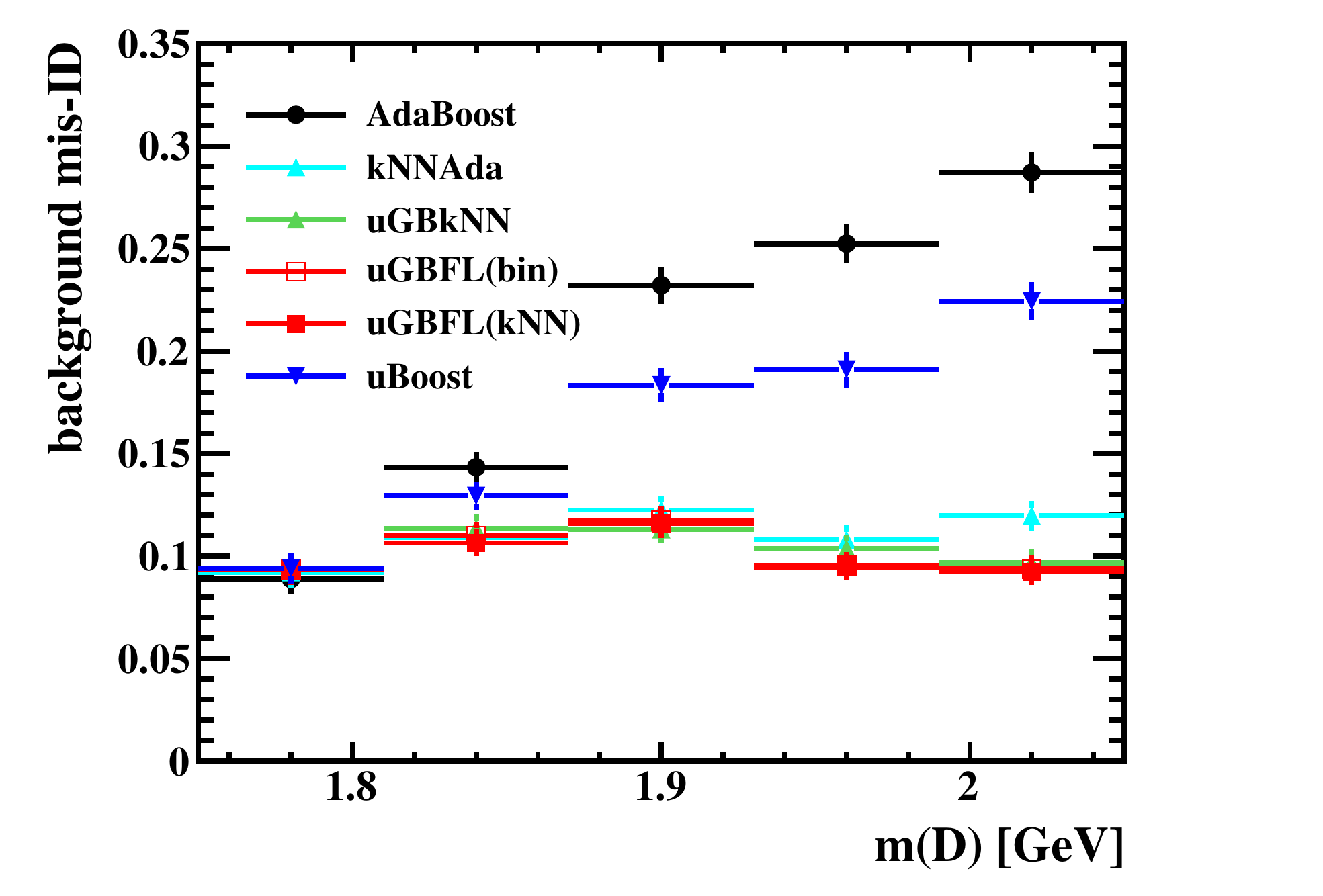}
  \caption{\label{fig:md_results} (left) Signal and (filled) background samples used in training.  (right) Background mis-identification {\em vs} $D$ candidate mass for an arbitrary working point of 10\% background mis-identification in the training region $1.75 < m(D) < 1.85$~GeV is displayed.}
\end{figure}

As a separate study using the same data samples, consider the case where one has simulated signal events and uses data from a {\em nearby} region, a so-called sideband, for background.  This is a common situation in particle-physics analyses.  Figure~\ref{fig:md_results} shows the training samples used.  A major problem can arise in these situations as typically input variates to the BDT are correlated with the parent particle mass.  Therefore, the BDT may learn to reject the background in the training using the fact that the mass of the background and signal candidates is different.  This is just an artifact of how the background sample is obtained and will not be true for background candidates {\em under} the signal peak.  Figure~\ref{fig:md_results} shows the background mis-identification rate {\em vs} $D$ candidate mass.  AdaBoost has clearly learned to use this mis-match in signal and background candidate masses in the training.   The background in the region of the signal is about three times higher than one would expect from looking only at the sideband data.

Figure~\ref{fig:md_results} also shows the background mis-identification rate {\em vs} $D$ candidate mass for the various uniform classifiers where $y = m(D)$ and the choice is for uniformity in the background efficiency\footnote{The algorithms in this paper can easily be made uniform on signal, background or both.}.  
The uBoost algorithm does better than AdaBoost here but is still not optimal.  The way that uBoost achieves uniformity is not such that it can be trusted to work outside the region of training.  The algorithms presented in this paper each does well in achieving similar performance in the training and signal regions.  
Consider, {\em e.g.}, the uGBFL approach to achieving uniform selection efficiency.  In this case the training drives the BDT response itself to have the same PDF everywhere in the region $1.75 < m(D) < 1.85$~GeV (the training region).  
This does not guarantee that the BDT response is truly independent of $m(D)$ outside the training region, but does strongly suppress learning to use $m(D)$ and in this example results in the desired behavior.  
Finally, if both high and low $m(D)$ sidebands had been used, it is possible for a BDT to create a fake peak near the signal peak location.  The use of uGBFL greatly reduces the chances and possible size of such an effect.

\section{CPU Resources}

One drawback of the uBoost technique is that it has a high degree of computational complexity: 
while AdaBoost trains $M$ trees (a user-defined number), uBoost builds $100 \times M$ trees.  
The algorithms presented in this paper only build $M$ trees; however, the boosting involves some more complicated algorithms.  Training each of the $M$ trees scales as follows for $N$ training events:
\begin{itemize}
       \item kNNAdaBoost: $O(k \times N)$;
	\item uGBkNNknn: $O(k \times N)$ for $A_{\rm knn}$, and  
	$O( \text{\#nonzero elements in the matrix})$ for arbitrary matrix $A$;
	\item uGBFL(bin): $O(N \ln N)$;
	\item uGBFL(kNN): $O(N \ln N + N k \ln k) $.
\end{itemize}
In the example analysis studied in this paper, we find that the training time for these new algorithms is within a factor of two the same as AdaBoost.  The CPU-resource usage of these new algorithms is not prohibitive.

\section{Summary}

A number of novel boosting algorithms have been presented that consider uniformity of selection efficiency in a multivariate space in addition to mis-classifcation errors.  Of these, the uGBFL algorithm has the best performance on the example analyses studied in this paper.
This algorithm is expected to be useful in a wide-variety of analyses performed in particle physics.

\section{Source code}

The code for classifiers proposed in this article as well as for metrics of uniformity is publicly available at repository 
{\tt \href{https://github.com/anaderi/lhcb_trigger_ml}{https://github.com/anaderi/lhcb\_trigger\_ml}}.

\acknowledgments

These results were obtained using events generated with the official LHCb simulation, and we are grateful to the LHCb collaboration for this privilege. 
We particularly acknowledge the work of the LHCb Simulation and Core Computing teams in tuning the simulation software and managing the productions of simulated events.
MW is supported by NSF grant PHY-1306550.

\appendix
\section{Measures of uniformity}
In this section we discuss different methods for measuring the uniformity of prediction. 
One typical way of 'checking' uniformity of prediction used by physicists 
is fitting the distribution of the events that were classified as signal (or background) over the 
feature for which you wish to check uniformity.
This approach requires assumptions about the shape of the distribution,
which makes quantitative comparisons of different classifiers difficult.
Our aim here is to explore uniformity figures of merit which make comparing classifiers easier,
analogously to how the area under the ROC curve can be used to compare absolute classifier performance.

The output of event classification is the probability of each event being signal or background,
and it is only after we apply a cut on this probability that events are classified.
An ideal uniformity of signal prediction can then be defined for a given ``uniform feature'' of interest.
It means that whichever cut we select,
the efficiency for a signal event to pass the cut doesn't depend on the uniform feature.
Uniformity for background can be defined in the same manner, but for simplicity,
in what follows we will only discuss the uniformity of efficiency for signal events.

A trivial example of a classifier that has ideal uniformity is a classifier which returns a random classification probability,
but such a classifier is of course not very useful. One can try to design a uniform classifier with respect to
a given feature by not using this feature, or any correlated features, in the classification; in practice, however,
this approach also tends to lead to poorly performing classifiers. The approach which we take in this paper is 
to explicitly let the classifier learn how to balance non-uniformities coming from different features in such a way
as to generate a classification which is uniform on average. It is then important to be able to accurately measure
the uniformity of classification. 

Before proceeding, it is useful to define some desirable properties of uniformity metrics
\begin{enumerate}
\item
The metric shouldn't depend strongly on the number of events used to test uniformity;
\item
The metric shouldn't depend on the normalization of the event weights: if we multiply all the weights by some arbitrary number, it shouldn't change at all;
\item 
The metric should depend only on the order of predictions, not the exact values of probabilities.
This is because we care about which events pass the cut and which don't, not about the exact values of predictions.
For example: correlation of prediction and mass doesn't satisfy this restriction.
\item
The metric should be stable against any of its own free parameters: if it uses bins, changing the number of bins shouldn't affect the result,
if it uses $k$-nearest neighbors, it should be stable against different values of $k$.
\end{enumerate}
In what follows we will consider different metrics which satisfy these criteria, and then compare their performance in
some test cases.

\subsection*{Standard Deviation of Efficiency on Bins (SDE)}

\def\bineff{\text{eff}_\text{bin}}
\def\binweight{\text{weight}_\text{bin}}
\def\globaleff{\text{eff}}
\def\SDE{\text{SDE}}
\def\bin{\text{bin}}

If the space of uniform features is split into bins, it is possible to define the global efficiency
\[
	\globaleff = \dfrac{
		\text{total weight of signal events that passed the cut}}
		{\text{total weight of signal events}},
\]

as well as the efficiency in every bin, 
\[
	\bineff = \dfrac{
		\text{weight of signal events in bin that passed the cut}}
		{\text{weight of signal events in this bin}}.
\]
One measure of non-uniformity is the standard deviation of bin efficiencies from the global efficiency:
\[
	\sqrt{\sum_{\bin} \left( \bineff - \globaleff \right)^2  }.
\]

To make the metric more stable against fluctuations in bins which contain very few events, we add weights to the bins (note that $\sum_\bin \binweight = 1$):
\[
	\binweight = \dfrac{\text{total weight of signal events in bin}}
		{\text{total weight of signal events}},
\]
giving the weighted standard deviation (SDE) formula
\[
	\SDE(\globaleff) = 
	\sqrt{\sum_{\bin} \binweight \times \left(\bineff - \globaleff \right)^2}. 
\] 


This formula is valid for any given cut value. To measure the overall non-flatness of the selection, we
take several global efficiencies and use
\[
	\SDE^2  =  \frac{1}{k} 
	\sum_{\globaleff \in [\globaleff_1 \dots \globaleff_k] }  
		\text{SDE}^2(\globaleff)
\]
Another power $p \neq 2$ can be used as well, but $p=2$ is considered as the default value.
%
%

\subsection*{Theil Index of Efficiency}
\def\theil{\text{Theil}}

The Theil Index is frequently used to measure economic inequality:
\[
	\theil = \frac{1}{N} \sum_i \frac{x_i}{<x>} \ln{\frac{x_i}{<x>}}, 
		\qquad <x> = \frac{1}{N} \sum_i x_i
\]
In our case we have to alter formula a bit to take into account that different bins have different impact, thus the formula turns into
\[
	\theil(\globaleff) = \sum_\bin \binweight \; \frac{\bineff}{\globaleff} \; \ln{\frac{\bineff}{\globaleff}}.
\]
To measure the overall non-flatness, we
average values for several global efficiencies:
\[
	\theil  =  \frac{1}{k} 
	\sum_{\globaleff \in [\globaleff_1 \dots \globaleff_k] }  
		\theil(\globaleff)
\]

\subsection*{Distribution Similarity Approach}
\label{sec:similarity}

\begin{figure}[h]
		\centering
		\begin{subfigure}[b]{0.48\textwidth}
			\includegraphics[width=\textwidth]{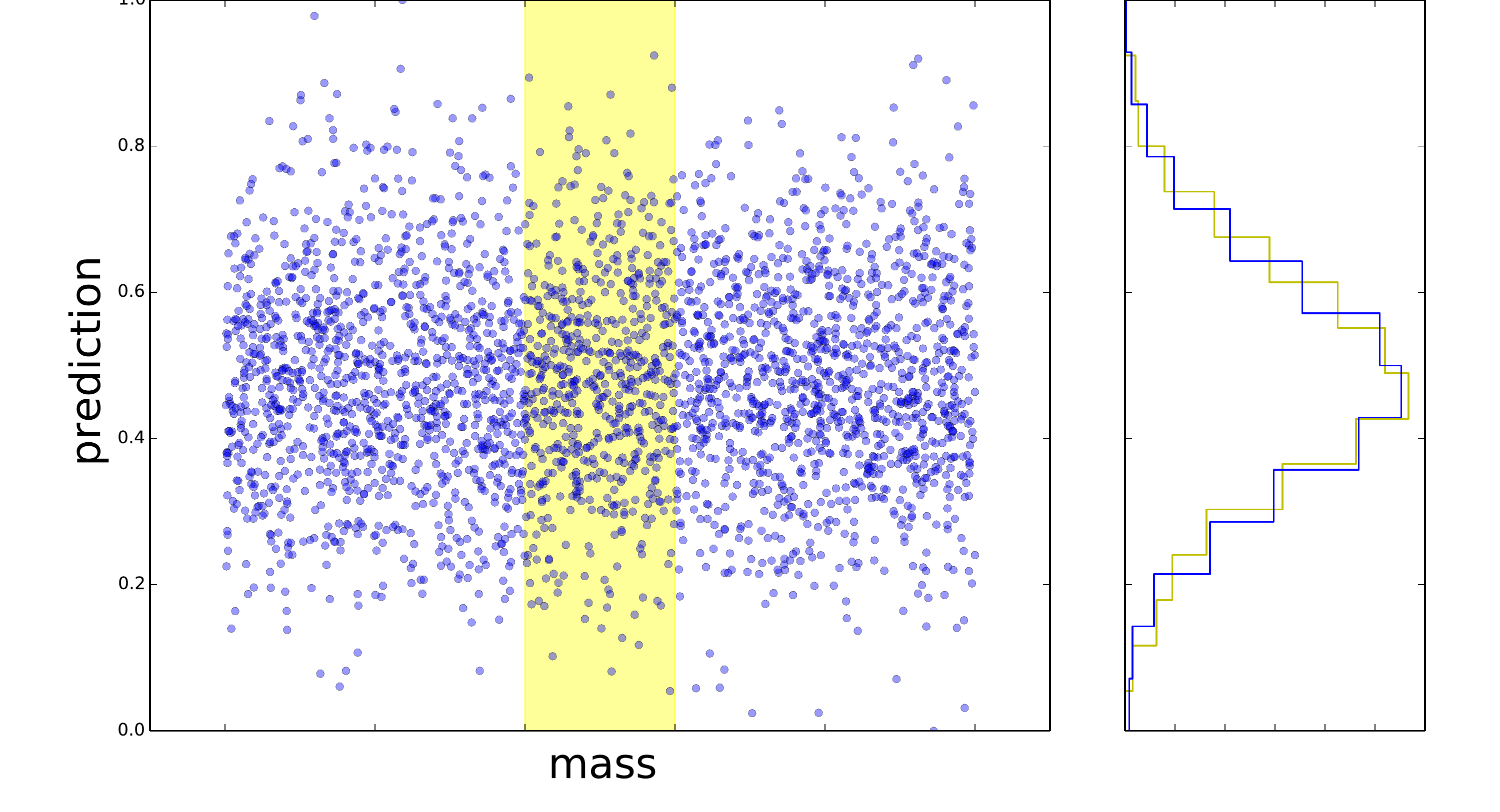}
		\end{subfigure}
		\begin{subfigure}[b]{0.48\textwidth}
			\includegraphics[width=\textwidth]{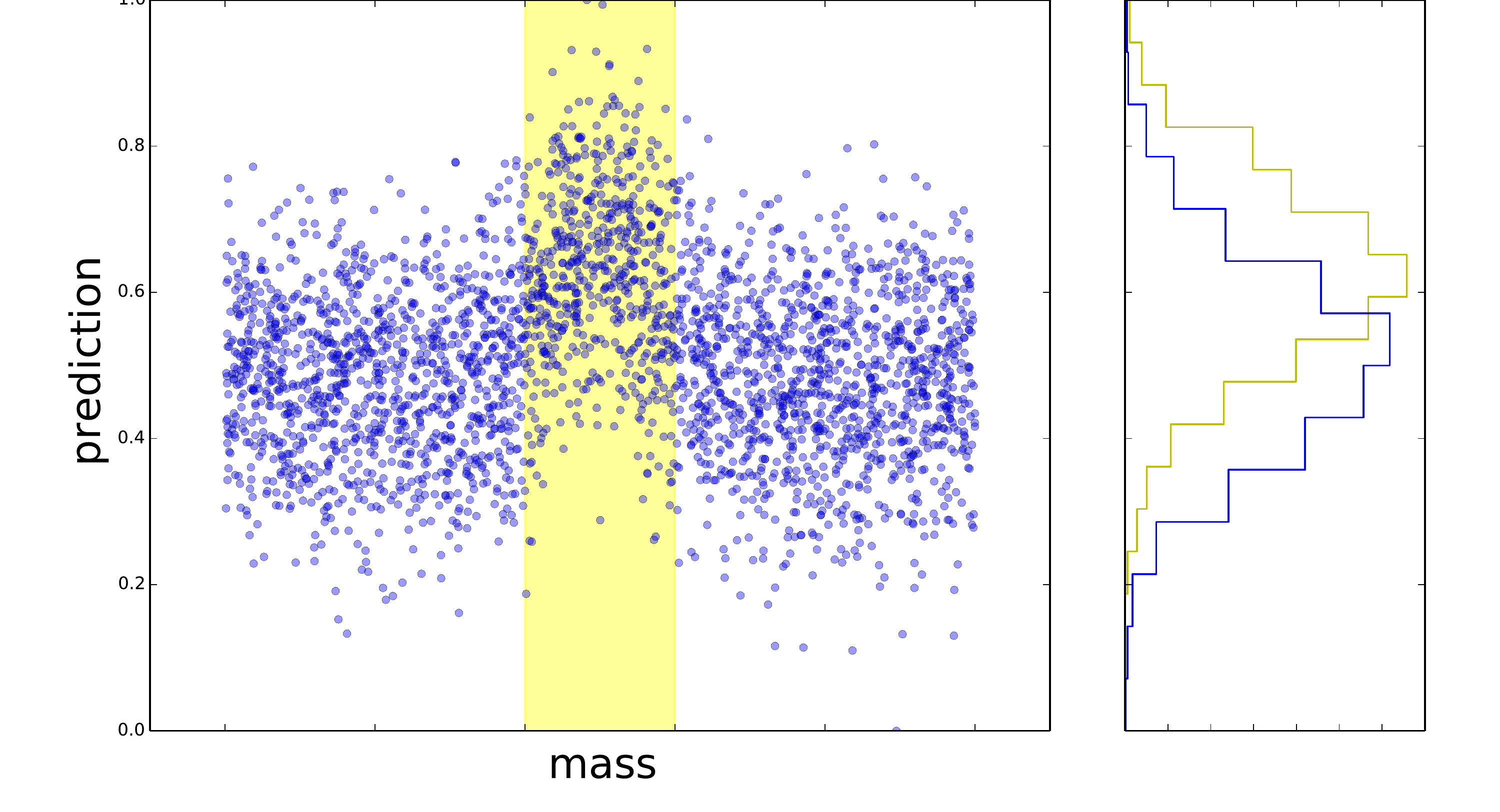}
		\end{subfigure}
		\caption{Demonstration of the distribution similarity approach. (left) Predictions are uniform in mass, the distribution of predictions in the bin (yellow) is close to the global (blue). (right) Distribution with peak in the middle, the distribution in the bin is quite different from the global distribution. In both cases the yellow rectangle shows the events in the bin over mass.\label{fig:dsavisualization}}
\end{figure}
Instead of measuring uniformity in terms of binned efficiencies, it is possible to consider the distribution of
the binned classifier predictions, $F_\bin$, directly.
Ideal uniformity means that all the distributions $F_\bin$ are equal and hence equal to the global distribution $F(x)$. 
This is demonstrated on figure \ref{fig:dsavisualization}.
To 'measure' non-flatness we can use some distribution distance, like Kolmogorov-Smirnov:
\[
	 \sum_{\bin} \binweight \max_x \abs{F_{\bin}(x) - F(x)},
\]
but Cram\'er--von Mises similarity is more informative (usually $p=2$ is used):
\[
	 \sum_{\bin} \binweight \int \abs{F_{\bin}(x) - F(x)}^p dF(x),
\]
in particular because Kolmogorov-Smirnov measures are too sensitive to local non-uniformities.
The advantage of this method is that we don't need to select some global efficiencies like in the previous metrics.
%
%

\subsection*{Knn-based modifications}

\def\knni{\text{knn}(i)}
\def\effknni{\text{eff}_{\knni}}
\def\weightknni{\text{weight}_{\knni}}
\def\Fknn{F_{\knni}}

\def\knnSDE{\text{knnSDE}}
\newcommand*\mean[1]{\overline{#1}}

Though operating with bins is usually both simple and very efficient, 
in many cases it is hard to find the optimal size of bins in the space of uniform features (specifically in the case of more than two dimensions).
As mentioned earlier, problems can also arise due to bins with very low populations.

In these cases we can switch to $k$-nearest neighbors: for each signal event we find $k$ nearest signal events (including the event itself)
in the space of uniform features. Now we can compute the efficiency $\effknni$, from the empirical distribution $\Fknn$ of nearest neighbors. 
The weights for $\knni$ are proportional to the total weight of events in $\knni$:
\[
	\weightknni = \alpha \sum_{j \in \knni} w_j, \qquad \alpha^{-1} = \sum_i \sum_{j \in \knni} w_j,
\]
so again weights are normed to 1: $\sum_{i} \weightknni = 1$. 

It is then possible to write the knn versions of SDE
\[
	\knnSDE^2(\globaleff)
		= \sum_{i \in \text{events}} \weightknni \abs{\effknni - \globaleff}^2,
\]
\[
	\knnSDE^2 = \frac{1}{k} \sum_{\globaleff \in [\globaleff_1, \dots \globaleff_k]} \knnSDE^2(\globaleff),
\]
the Theil index of efficiency
\[
	\text{knnTheil}(\globaleff) = \sum_{i \in \text{events}} \weightknni \; \frac{\effknni}{\globaleff} \; \ln{\frac{\effknni}{\globaleff}},
\]
\[
	\text{knnTheil} = \frac{1}{k} \sum_{\globaleff \in [\globaleff_1, \dots \globaleff_k]} \text{knnTheil}(\globaleff),
\]
and the similarity-based measure:
\[
	 \sum_{i \in \text{events}} \weightknni \int \abs{\Fknn(x) - F(x)}^p dF(x).
\]

The knn approach suffers from a drawback: the impact of different events has very little connection with the weights,
because some events are selected as nearest neighbours much more frequently than others.
This effect can be suppressed by dividing the initial weight of the event by the number of times it is selected 
as a nearest neighbour. 

\subsection*{Advantages and Disadvantages of Different Metrics}
\subsubsection*{Theil and SDE}
Let us compare two metrics that have proven most appropriate for our problem, SDE and Theil. We have some masses distributed uniformly in $[0,1]$, some constant $\alpha$ from interval $[0,1]$. 
The predictions are correlated with mass via beta distribution.
First distribution with a peak is obtained by generating a prediction for each event according to its mass:
\[
	p_{\rm peak} \thicksim \text{Beta}(1 + \alpha f(m), 1), \qquad  f(m) = 5 \times e^{- 100  \abs{m - \overline{m}}^2}
\]
The second distribution is obtained by flipping:
\[
	p_{\rm pit} \thicksim 1 - \text{Beta}(1 + \alpha f(m), 1)
\]
SDE should show no changes if we flip the distribution, while the Theil should make difference between pit and peak. 

\begin{figure}[h]
\centering
		\begin{subfigure}[b]{0.95\textwidth}
			\includegraphics[width=\textwidth]{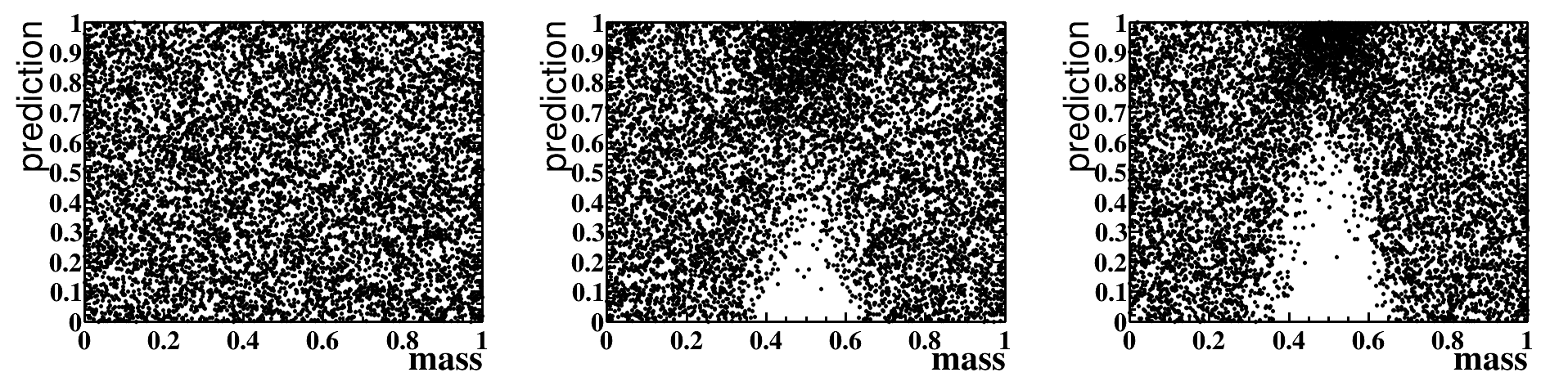}
			\caption{Mass vs prediction.}
		\end{subfigure}
		\begin{subfigure}[b]{0.95\textwidth}
			\includegraphics[width=\textwidth]{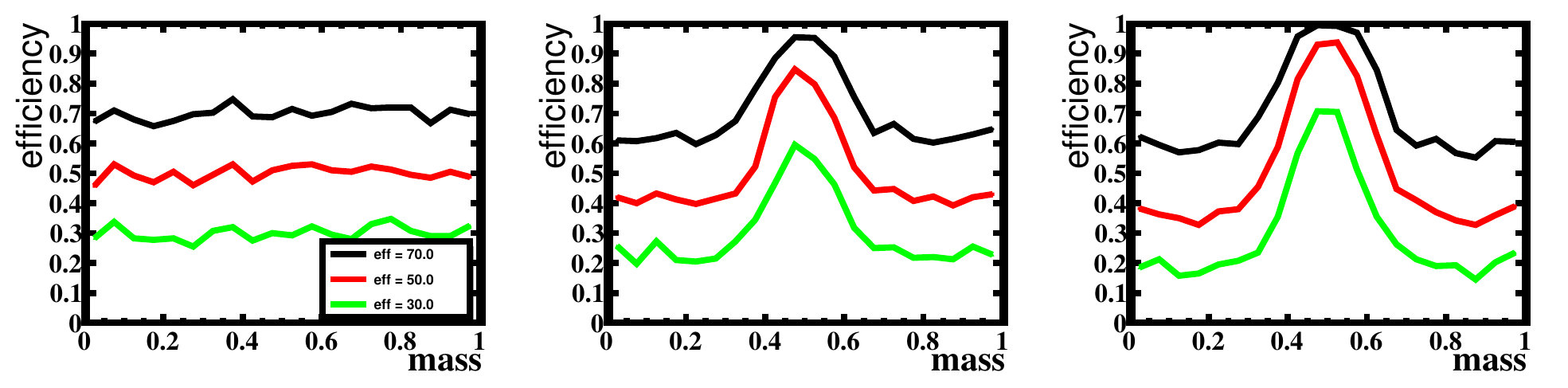}
			\caption{Mass vs efficiency. The different coloured lines correspond to
				different global classifier efficiencies, as explained in the legend on the leftmost subplot.}
		\end{subfigure}
		\caption{Peak response distribution and efficiencies as a function of mass. \label{fig:peakdists}}
\end{figure}

\begin{figure}[h]
\centering
		\begin{subfigure}[b]{0.95\textwidth}
			\includegraphics[width=\textwidth]{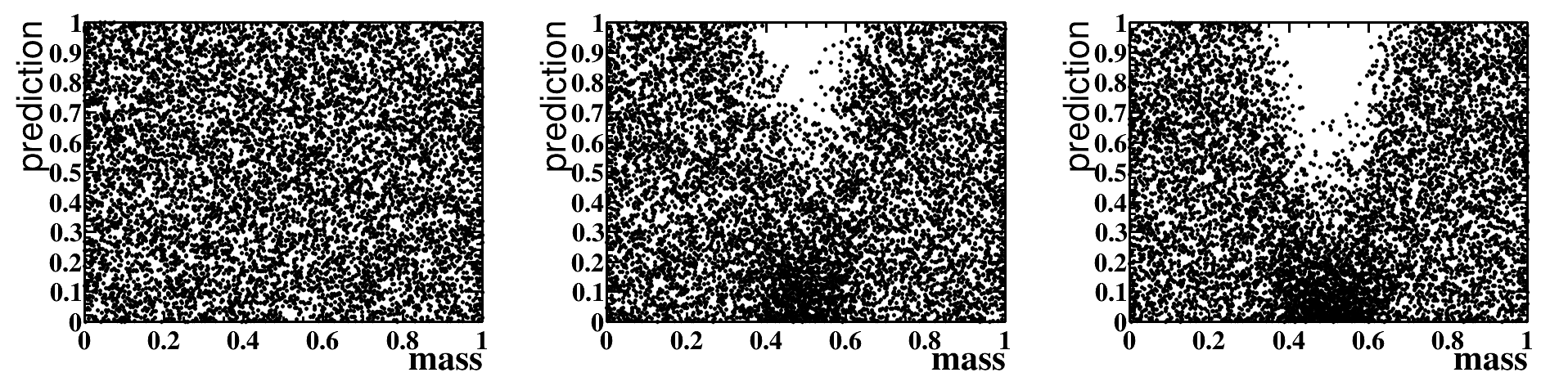}
			\caption{Mass vs prediction.}
		\end{subfigure}
		\begin{subfigure}[b]{0.95\textwidth}
			\includegraphics[width=\textwidth]{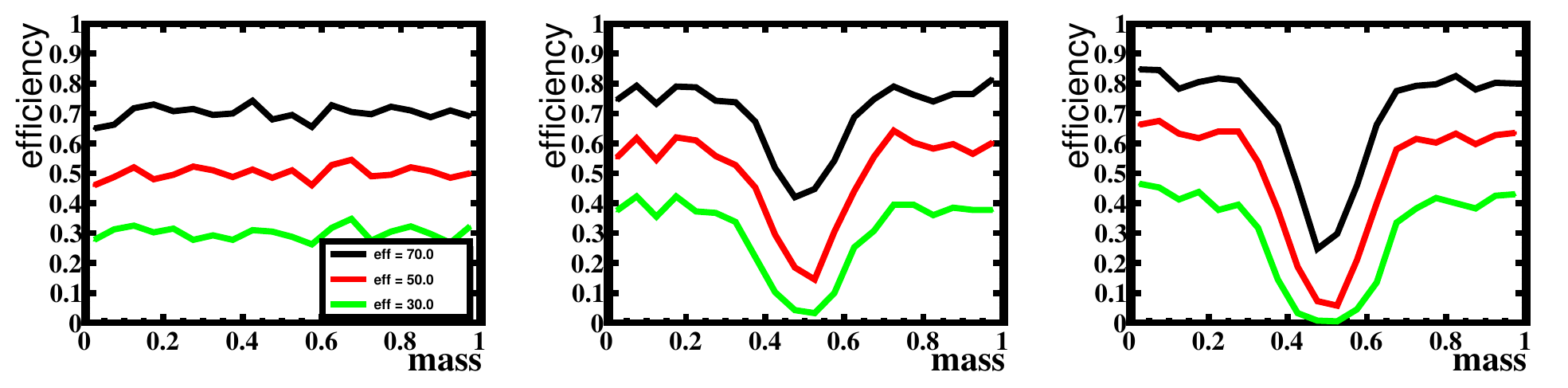}
			\caption{Mass vs efficiency. The different coloured lines correspond to
				different global classifier efficiencies, as explained in the legend on the leftmost subplot.}
		\end{subfigure}
		\caption{Pit response distribution and efficiencies as a function of mass. \label{fig:pitdists}}
\end{figure}

Figures~\ref{fig:peakdists} and~\ref{fig:pitdists} show the distribution of the predictions
and classifier efficiency as a function of mass for the peak and pit distributions, respectively. 

\begin{figure}[h]
\centering
	\includegraphics[width=0.9\textwidth]{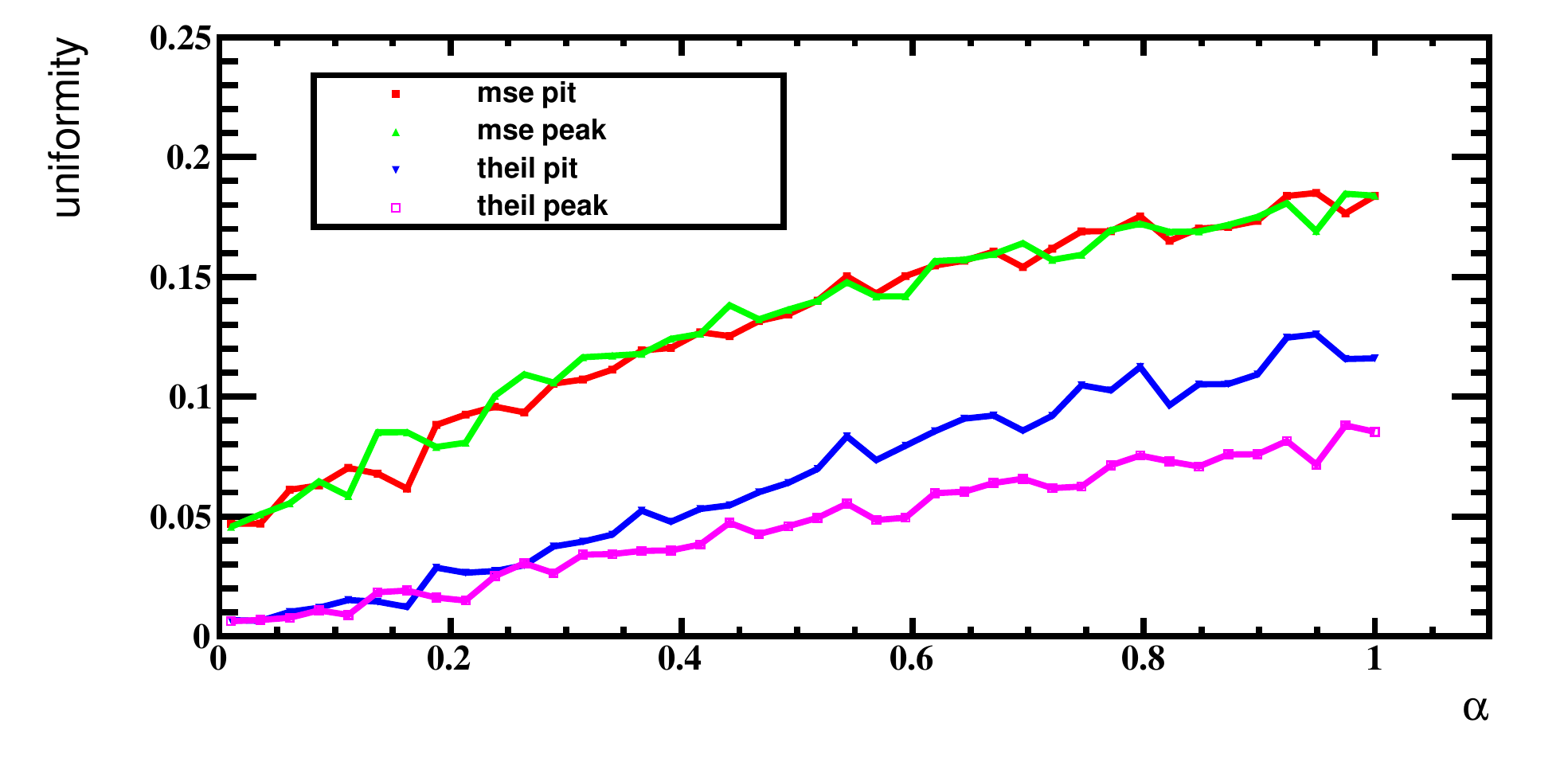}
	\caption{Comparison of SDE and Theil uniformity measures for different values of the constant $\alpha$ for the peak and pit distributions.
The Theil can clearly distinguish between these two while the SDE measure cannot.\label{fig:pitvspeak}}
\end{figure}

From figure~\ref{fig:pitvspeak} we can see that SDE doesn't make any difference between these distributions, while Theil has lower value in the second case 
which indicates that distribution is flatter. This example demonstrates that Theil has larger penalty for distributions with narrow peaks rather than with narrow pits in the distribution of efficiecies.

\subsubsection*{$D \to hhh$}
\begin{figure}[h]
\centering
		\begin{subfigure}[b]{0.48\textwidth}
			\includegraphics[width=\textwidth]{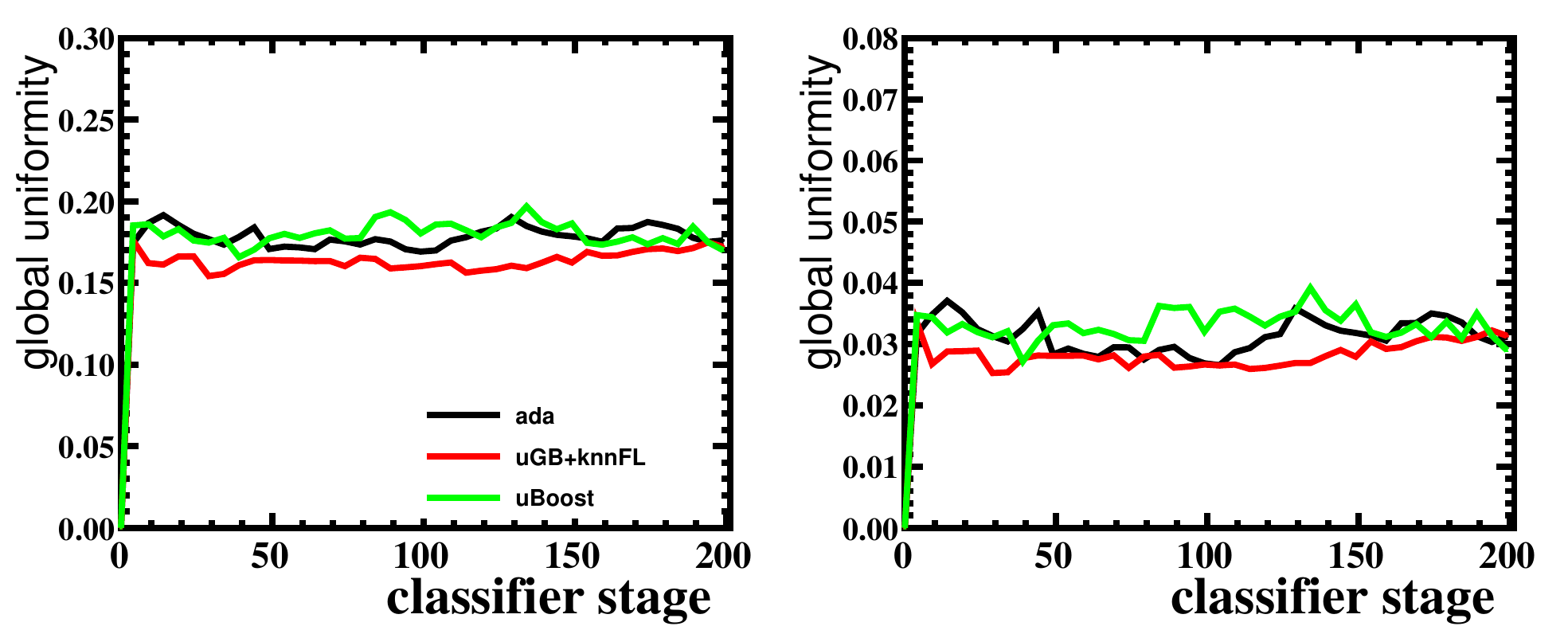}
			\caption{Metrics for $D \to hhh$ signal.}
		\end{subfigure}
		\begin{subfigure}[b]{0.48\textwidth}
			\includegraphics[width=\textwidth]{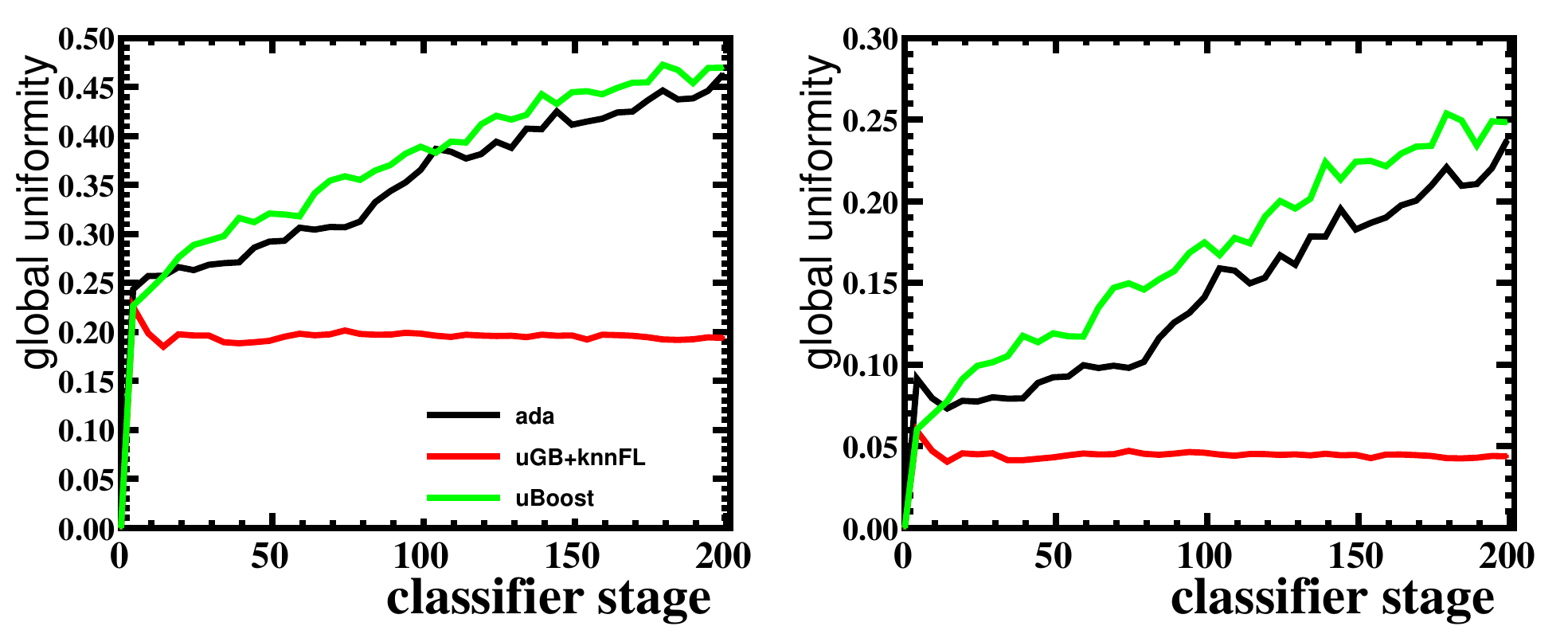}
			\caption{Metrics for $D \to hhh$ background.}
		\end{subfigure}
		\caption{Uniformity metrics for the $D \to hhh$ data set. The (top) signal and (bottom) background uniformities are plotted as a function of the training stage of a given classifier, listed in the legends on the leftmost plots. Within each figure the leftmost plot is the SDE and the rightmost plot is the Theil uniformity metric. \label{fig:d2hhhsdetheil}}
\end{figure}

Finally, we compare the SDE and Theil uniformity measures for the classifiers applied to the $D \to hhh$ data set, as shown in Fig.~\ref{fig:d2hhhsdetheil}.
Both measures show similar results, so there is no significant difference between them, nor the CvM metric, on this dataset.

\begin{thebibliography}{9}

\bibitem{ref:lhcbhlt} 
R. Aaij {\em et al.} [LHCb Trigger Group], 
{\em The LHCb trigger and its performance}, 
JINST {\bf 8}, P04022 (2013). 
[arXiv:1211.3055]

\bibitem{ref:bbdt}  
V. V. Gligorov and M. Williams, 
{\em Efficient, reliable and fast high-level triggering using a bonsai boosted decision tree}, 
JINST {\bf 8}, P02013 (2013). 
[arXiv:1210.6861]

\bibitem{ref:FS1997}
{Y. Freund, and R. Schapire},
\emph{A Decision-Theoretic Generalization of On-Line Learning and an Application to Boosting}, 
{\emph{Journal of Computer and System Sciences} {\bf 55(1)} (August 1997) 119-139}.

\bibitem{ref:uboost} 
J. Stevens and M. Williams,
{\em uBoost: A boosting method for producing uniform selection efficiencies from multivariate classifiers}, 
JINST {\bf 8}, P12013 (2013). 
[arXiv:1305.7248]

\bibitem{ref:F1999}
{Jerome H. Friedman},
\emph{Greedy Function Approximation: A Gradient Boosting Machine},
\href{http://statweb.stanford.edu/~jhf/ftp/trebst.pdf}
{\emph{Annals of Statistics} {\bf 29} (2000) 1189-1232}.



\bibitem{Sjostrand:2006za}
T.~Sj\"{o}strand, S.~Mrenna and P.~Skands,
\emph{PYTHIA 6.4 physics and manual},
\href{http://dx.doi.org/10.1088/1126-6708/2006/05/026}
{\emph{JHEP} {\bf 05} (2006) 026}.
[\hepph{0603175}]

\bibitem{LHCb-PROC-2010-056}
I.~Belyaev, I. {\em et al.},
\emph{Handling of the generation of primary events in GAUSS, the LHCb simulation framework},
\href{http://dx.doi.org/10.1109/NSSMIC.2010.5873949}
{\emph{NSS/MIC} {\bf IEEE} (2010) 1155}.

\bibitem{Lange:2001uf}
D.J.~Lange,
\emph{The EvtGen particle decay simulation package},
\href{http://dx.doi.org/10.1016/S0168-9002(01)00089-4}
{\emph{NIM} {\bf A462} (2001) 152-155}.

\bibitem{Golonka:2005pn}
P.~Golonka and Z.~Was,
\emph{PHOTOS Monte Carlo: a precision tool for QED corrections in $Z$ and $W$ decays},
\href{http://dx.doi.org/10.1140/epjc/s2005-02396-4}
{\emph{EPJ} {\bf C45} (2006) 97-107}.
[\hepph{0506026}]

\bibitem{Allison:2006ve}
J.~Allison {\em et al.},
\emph{Geant4 developments and applications},
\href{http://dx.doi.org/10.1109/TNS.2006.869826}
{\emph{IEEE TNS} {\bf 53} (2006) 270}.

\bibitem{Agostinelli:2002hh}
S.~Agostinelli {\em et al.},
\emph{Geant4: a simulation toolkit},
\href{http://dx.doi.org/10.1016/S0168-9002(03)01368-8}
{\emph{NIM} {\bf A506} (2003) 250}.

\bibitem{LHCb-PROC-2011-006}
M~Clemencic {\em et al.},
\emph{The LHCb simulation application, GAUSS : design, evolution and experience},
\href{http://dx.doi.org/10.1088/1742-6596/331/3/032023}
{\emph{J. Phys. Conf. Ser.} {\bf 331} (2011) 032023}.










\end{thebibliography}
\end{document}